\documentclass[final,5p,times,twocolumn,a4paper]{elsarticle}

\usepackage{amssymb}
\usepackage{amsmath}

\usepackage{lipsum}

\usepackage{tikz}
\usetikzlibrary{shapes,arrows}
\usetikzlibrary{patterns}

\usepackage{pgfplots}
\pgfplotsset{compat=1.13}
\usepgfplotslibrary{fillbetween}

\definecolor{PrimalOrange}{RGB}{255,153,85}
\definecolor{DualBlue}{RGB}{0,255,255}
\definecolor{SlabGreen}{RGB}{170,212,0}

\usepackage{hyperref}
\usepackage{xcolor}

\newcommand{\doi}[1]{\href{https://doi.org/#1}{doi:#1}}
\newcommand{\urn}[1]{\href{http://nbn-resolving.de/#1}{#1}}
\newcommand{\arXiv}[1]{\href{https://arxiv.org/abs/#1}{arXiv:#1}}

\journal{SoftwareX}

\begin{document}

\begin{frontmatter}



\title{Efficient and scalable data structures and algorithms for\\
goal-oriented adaptivity of space-time FEM codes}


\author[hsuhh1]{Uwe K\"ocher$^{*,}$}
\author[hsuhh1]{Marius Paul Bruchh\"auser}
\author[hsuhh1]{Markus Bause}

\address[hsuhh1]{Helmut-Schmidt-University,
University of the Federal Armed Forces Hamburg,
Numerical Mathematics, Holstenhofweg 85, 22043 Hamburg\\
koecher@hsu-hamburg.de ($^*\!\!$ corresponding author),
\{bruchhaeuser,bause\}@hsu-hamburg.de}

\begin{abstract}
The cost- and memory-efficient numerical simulation of
coupled volume-based multi-physics problems
like flow, transport, wave propagation and others
remains a challenging task with finite element method (FEM) approaches.
Goal-oriented space and time adaptive methods derived from the
dual weighted residual (DWR) method appear to be a shiny key technology to
generate optimal space-time meshes to minimise costs.
Current implementations for challenging problems of numerical screening tools
including the DWR technology broadly suffer in their extensibility to other problems,
in high memory consumption or in missing system solver technologies.
This work contributes to
the efficient embedding of DWR space-time adaptive methods
into numerical screening tools
for challenging problems of physically relevance
with a new approach of flexible data structures and algorithms on them,
a modularised and complete implementation as well as
illustrative examples to show the performance and efficiency.

\vskip54ex
~
\end{abstract}


\end{frontmatter}


\section{Motivation and significance}
\label{sec:1}

\begin{figure*}[t!b]
\centering

\includegraphics[height=5.0cm]{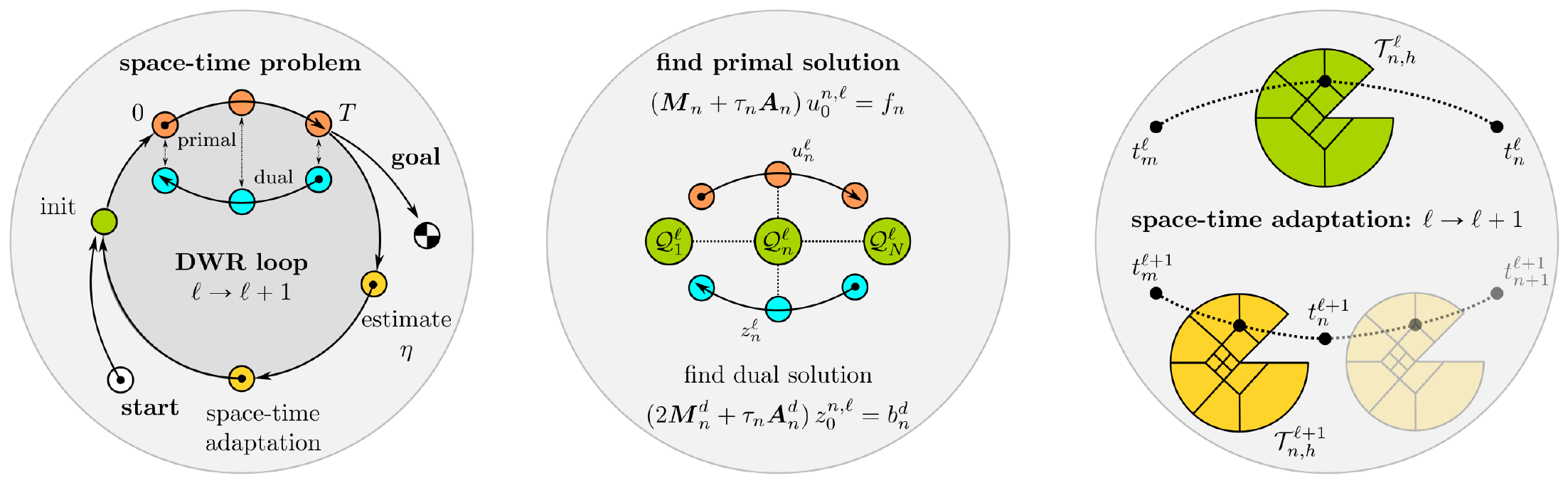}
\caption{General workflow of the space-time adaptive numerical simulation tools
\texttt{DTM++.Project}/\texttt{dwr}-*
using the dual weighted residuals (DWR) technique for the error estimation (left),
an exemplary forward and backward time marching step $n$ on $(d$+$1)$-dimensional
space-time slabs $\mathcal{Q}_n^\ell$ in the DWR loop $\ell$ (middle) and
an exemplary space-time adaptation update of the spatial triangulation
$\mathcal{T}_{h,n}^\ell \to \mathcal{T}_{h,n}^{\ell+1}$
and the temporal subinterval $I_n^\ell \to I_n^{\ell+1}$
to prepare the DWR loop $\ell$+$1$ (right).}
\label{fig:1:space-time-dwr}
\end{figure*}

\subsection{Introduction}
\label{sec:1:1:introduction}

The accurate, reliable and efficient numerical approximation of flow in
heterogeneous deformable porous media
include several coupled multi-physics phenomena,
such as multi-phase diffusion and convection-dominated transport
with underlying chemical reactions,
deformation and poroelastic wave propagation
as well as fluid-structure interactions in a more general sense,
and is of fundamental importance in
environmental, civil, energy, biomedical and many other engineering fields
to yield a cost-effective numerical simulation screening tool to support
the still challenging and fundamental research of understanding
such multi-physics phenomena.
Considerable representatives are coupled problems based on
the Navier-Stokes equations for incompressible flow,
the Euler equations for compressible flow,
the convection-diffusion-reaction equations and
the Biot-Allard equations for coupled flow with poroelastic wave propagation,
which are characterised by partial differential equations of
nonlinear and instationary character.
Space-time finite element method (FEM) approximations
offer appreciable benefits over finite difference and finite volume methods
such as the flexibility with which they can accommodate
discontinuities in the model, material parameters and boundary conditions
as well as for a priori and a posteriori error estimation
to establish optimal computational space-time grids in a self-adaptive way.
The dual weighted residual (DWR) method for goal-oriented adaptivity was
introduced by Becker and Rannacher (\cite{Becker1998,Becker2001})
and further studied;
cf. \cite{Endtmayer2017,Bruchhaeuser2018,Bruchhaeuser2018a,Bangerth2003}
and references therein.
The DWR method facilitates to find
optimal spatial mesh adaptations,
optimal temporal mesh adaptations,
as well as optimal
  local space-time polynomial degrees,
  tuning parameters and
  physical or numerical models,
such that the overall computational costs are minimised
for reaching a goal in a target quantity of interest.

A target quantity is a cost, error or energy functional
$\mathcal J$ and depends on an user-chosen quantity of interest.
Examples for a quantity of interest are
the stress derived from the primal displacement variables or
the drag coefficient derived from the primal fluid velocity variables
of the underlying problem.
The goal of the DWR method is to satisfy a guaranteed error bound
in the target quantity of interest
instead of satisfying a classical error bound of
the primal solution variable in a standard norm.

Unfortunately, the DWR method still has challenging drawbacks compared to
standard a posteriori error based adaptive methods.
It needs to solve the primal or forward problem and
  an auxiliary dual or adjoint problem of higher approximation quality
  in each loop of an optimisation problem,
it needs variational space-time discretisations for the primal and dual problem,
  which yield problems on $(d$+$1)$-dimensional domains,
and there is a lack in efficient data structures, algorithms as well as
(non-)linear system solver and preconditioning technologies
  for sophisticated problems of physical interest,
which will be at least partially resolved by this work.

Fortunately, the DWR method works in general situations,
in which the problem itself and the quantity of interest are of nonlinear fashion,
and it does not rely on generally unknown assumptions of
the initial space-time mesh.
The key idea of the DWR method is to embed the given problem
  of finding solutions of a partial differential equation
into the framework of optimal control
to estimate the functional error of $\mathcal{J}(u) - \mathcal{J}(u_{\tau,h})$,
derive computable a posteriori error estimates $\eta$ with an approximation of
the corresponding dual solution $z$ of an auxiliary problem,
execute space-time mesh (and other) adaptations based on the information given by
$\eta$ and loop until the goal in the target quantity is reached
as it is outlined by Fig. \ref{fig:1:space-time-dwr}.
Consider to solve the constrained optimisation problem given by
\begin{equation}
\label{eq:1:dwr}
\mathcal{J}(u) = \min!\,,
\end{equation}
for a (cost, error or energy) functional $\mathcal{J}$ depending on $u$,
under the constraint of finding $u \in V$ from the variational problem
$$
0 = \mathcal{F}(\varphi) - \mathcal{A}(u)(\varphi)\,,\,\,
\text{for all test functions } \varphi \in W\,,
$$
which is set up from the partial differential equation problem
in a standard way.
$V$ and $W$ are corresponding variational trial and test space-time functional
spaces such that primal and dual solutions exist at least locally,
are unique and stable, but the latter ones can not be guaranteed for
arbitrary nonlinear problems without restrictions.
The calculus of variations theory associates
a corresponding Lagrangian functional $\mathcal{L}$ to \eqref{eq:1:dwr},
\begin{equation}
\label{eq:2:Lagrangian}
\mathcal{L}(u,z) := \mathcal{J}(u) + \mathcal{F}(z) - \mathcal{A}(u)(z)\,,
\end{equation}
with $z \in W$ as a Lagrange multiplier or, equivalently,
as the adjoint or dual solution variable,
for the global optimisation problem \eqref{eq:1:dwr}.
The primal solution $u$ is the first component of a stationary point
$(u,z)$ of $\mathcal{L}$
whereas the second component yields the corresponding dual solution $z$;
cf. \cite{Becker1998,Becker2001,Bangerth2003} for details.
Due to the mathematically challenging theory of even local existence of solutions
in general nonlinear settings, we restrict ourselves further,
without the loss of generality and the impact of our software,
to a linear prototype model for instationary transport in heterogeneous porous media.

\subsection{Exact Scientific Problem solved by the Software}
\label{sec:1:2:ExactSciProblemSoftware}

The \texttt{DTM++.Project}/\texttt{dwr-diffusion} frontend simulation tool
for the established finite element analysis library \texttt{deal.II}
\cite{dealIIReferenceV90}
yields a reference implementation of the prototype model for instationary transport
in heterogeneous porous media.
We consider the goal-oriented approximation of a target quantity $\mathcal{J}(u)$
subject to find $u \in V$ from the diffusion equation
\begin{equation}
\label{eq:3:model}
\rho\, \partial_t u - \nabla \cdot (\epsilon\, \nabla u) = f\,,
\end{equation}
in the space-time domain $\mathcal{Q} = \Omega \times I$
with $\Omega \subset \mathbb{R}^d$ (open and bounded), 
dimension $d=2,3$, and $I=(t_0,T)$, $0 \le t_0 < T < \infty$,
and equipped with appropriate initial and boundary conditions
$$
\begin{array}{r@{\,}c@{\,}l@{\,}l@{\,}l}
u(\boldsymbol x, t_0) &=& u_0(\boldsymbol x) & \text{in} & \Omega \times \{t_0\}\,,\\[1ex]
u(\boldsymbol x, t) &=& g(\boldsymbol x, t) & \text{on} & \Gamma_D \times I\,,\\[1ex]
\epsilon(\boldsymbol x)\, \nabla u(\boldsymbol x, t) \cdot \boldsymbol n(\boldsymbol x)
  &=& h(\boldsymbol x, t) &
  \text{on} & \Gamma_N \times I\,,\\[1ex]
\end{array}
$$
with the partition of the boundary $\partial \Omega = \Gamma_D \cup \Gamma_N$
and $\Gamma_D \neq \emptyset$.

A target quantity $\mathcal{J}(u)$ of interest derived from the primal solution
$u$ of \eqref{eq:3:model} might represent a local pollution concentration
under diffusive transport.
The coefficient functions $\rho > 0$ and $\epsilon > 0$ are
the mass density and the permeability of the medium, respectively,
and the right hand side function $f : \mathcal{Q} \to \mathbb{R}$ represents
volume forces acting on the interior domain $\Omega$ such as source or sink terms.
In \eqref{eq:3:model}, the differential operators $\partial_t(\cdot)$,
$\nabla \cdot (\cdot) = \partial_{x_1}(\cdot) + \dots + \partial_{x_d} (\cdot)$
and $\nabla(\cdot) = [ \partial_{x_1}(\cdot), \dots, \partial_{x_d}(\cdot) ]^T$
denote
the partial derivative for the time variable $t$,
the divergence and the gradient for the space variables $x_1,\dots,x_d$,
respectively, and, $\boldsymbol n$ denotes the outward facing normal vector,
in standard notation.
The initial-boundary value problem is closed by the choice of an appropriate
initial value function $u_0 : \Omega \to \mathbb{R}$
and a boundary value function $g : \Gamma_D \times I \to \mathbb{R}$
acting on the Dirichlet part of the boundary partition $\Gamma_D \neq \emptyset$.
Additionally, there is the choice
for an appropriate inhomogeneous Neumann boundary type condition function
$h : \Gamma_N \times I \to \mathbb{R}$ if $\Gamma_N \neq \emptyset$.

The weak variational primal problem, for brevity consider to find
$u := \tilde u - g$ or put $g=0$ without loss of generality
of the following discussion, reads as:
Seek $u \in V$, $V = L^2(0,T;H^1_0(\Omega))$,
satisfying $u(0) = u_0$, $u_0 \in H^1_0(\Omega)$, such that
\begin{equation}
\label{eq:4:primal}
A(u)(\varphi) = F(\varphi)\,,\,\,
\forall \varphi \in W=L^2(0,T;H^1_0(\Omega))\,,
\end{equation}
with the left-hand side bilinear form $A : V \times W \to \mathbb{R}$,
$$
A(u)(\varphi) := \displaystyle \int_0^T \int_\Omega \Big(
\varphi \cdot \rho\, \partial_t u
+ \nabla \varphi \cdot (\epsilon\, \nabla u)
\Big)\, \operatorname{d} \boldsymbol x\, \operatorname{d} t\,,
$$
and the right-hand side linear form $F : W \to \mathbb{R}$,
$$
F(\varphi) := \displaystyle \int_0^T \int_\Omega
\varphi \cdot f\, \operatorname{d} \boldsymbol x\, \operatorname{d} t\,,\,\,
f \in L^2(0,T; L^2(\Omega))\,.
$$
Remark that $\tilde u \in g + V$,
with $\tilde u$ having inhomogeneous Dirichlet boundary conditions $g \neq 0$,
can be generated in a standard way as it will be explained in the following
algorithm.
Briefly,
the weak variational dual problem results from the stationary condition of
$\mathcal{L}(u,z)$ \eqref{eq:2:Lagrangian} given by
\begin{equation}
\label{eq:5:stationary}
\Bigg\{\begin{array}{c}
\mathcal{J}(u+\psi)-\mathcal{J}(u) - A(\psi)(z)\\
F(\varphi) - A(u)(\varphi)
\end{array}\Bigg\}
\stackrel{!}{=}
\Bigg\{\begin{array}{c}
0\\
0
\end{array}\Bigg\}\,,
\end{equation}
for $\psi \in V$ and $z \in W$;
using the Euler-Lagrange method of constrained optimisation,
cf. for details \cite[Sec. 6.1]{Bangerth2003}, \cite{Bruchhaeuser2018}.
The dual problem for a general nonlinear goal functional $\mathcal{J}(u)$
reads as:
Find $z \in W$, $W=L^2(0,T;H^1_0(\Omega))$,
satisfying $z(T)=z_T$, $z_T \in H^1_0(\Omega)$,
such that
\begin{equation}
\label{eq:6:dual}
A^\ast(z)(\psi) = \mathcal{J}(u+\psi)-\mathcal{J}(u)
- \int_\Omega \psi(T) \cdot z(T)\, \operatorname{d} \boldsymbol x\,,
\end{equation}
for all $\psi \in V$, $V = L^2(0,T;H^1_0(\Omega))$, and
with the left-hand side adjoint bilinear form $A^\ast : W \times V \to \mathbb{R}$,
$$
A^\ast(z)(\psi) := \displaystyle \int_T^0 \int_\Omega \Big(
\psi \cdot \rho\, \partial_t z
- \nabla \psi \cdot (\epsilon\, \nabla z)
\Big)\, \operatorname{d} \boldsymbol x\, \operatorname{d} t\,,
$$
which is derived from integration by parts in time of $A(\psi)(z)$ of \eqref{eq:4:primal}
and put this into the first equation of \eqref{eq:5:stationary}.
Remark that the choice of trial and test functions for the dual problem here
yields traces on $t_0$ and $T$.
To estimate the functional error of $\mathcal{J}(u) - \mathcal{J}(u_{\tau,h})$
and to derive computable a posteriori error estimates $\eta$ for the space-time
mesh adaptivity, we need to discretise the primal problem \eqref{eq:4:primal}
and the dual problem \eqref{eq:6:dual}.

\begin{figure}[htb]
\centering

\includegraphics[width=.85\linewidth]{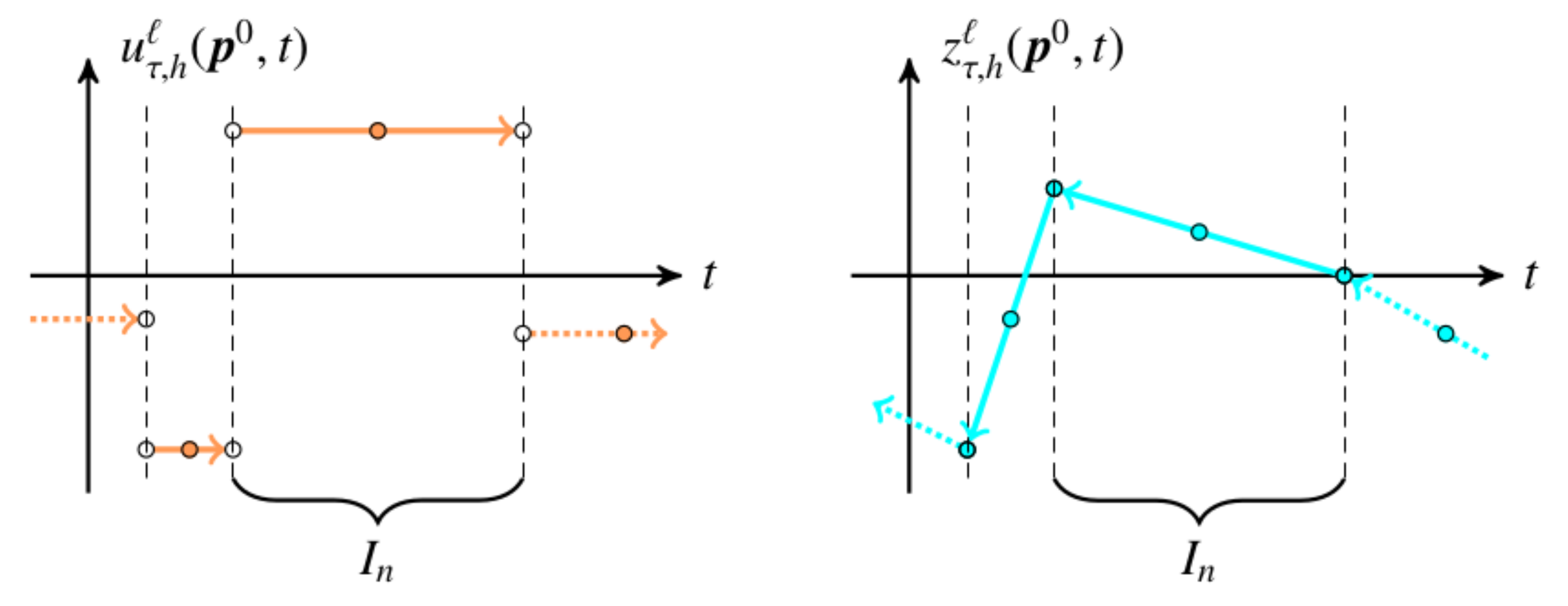}

\caption{Exemplary illustration of fully discrete primal and dual solutions
for a fixed observation point $\boldsymbol p^0 \in \Omega$ in space
in a DWR loop $\ell$ over time;
a piecewise constant discontinuous in time primal solution
$u_{\tau,h}^\ell(\boldsymbol p^0, t)$ of \eqref{eq:7:primal:fullydiscrete} (left),
and,
a piecewise linear continuous in time dual solution
$z_{\tau,h}^\ell(\boldsymbol p^0, t)$ of \eqref{eq:8:dual:fullydiscrete}
for a specific goal functional $\mathcal{J}(u)$ (right),
and the drawn arrows represent the implemented path of the natural time integration
on a ($d$+$1$)-slab approach.}
\label{fig:2:fullydiscrete}
\end{figure}

\begin{figure}[t]
\centering

\includegraphics[width=.85\linewidth]{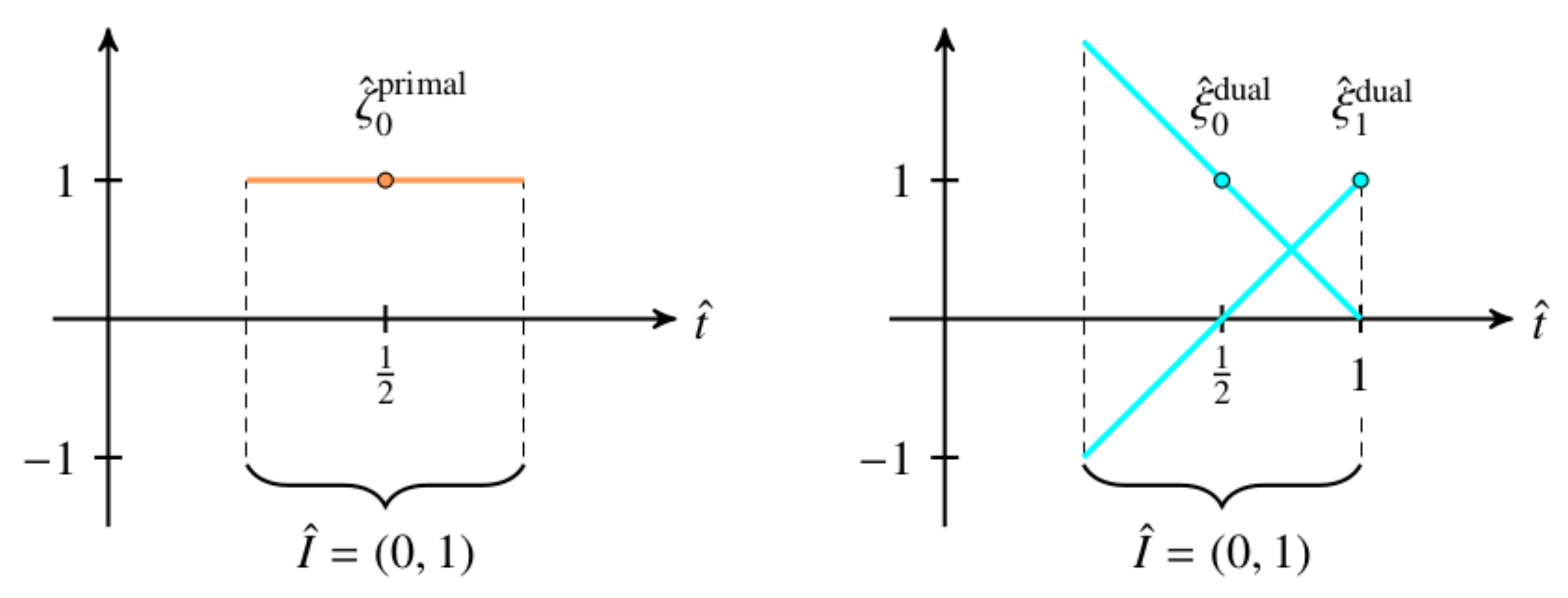}

\caption{Illustration of temporal reference trial basis functions
on the reference subinterval $\hat I = (0,1)$ 
for the primal solution \eqref{eq:7:primal:fullydiscrete} (left) and
the dual solution \eqref{eq:8:dual:fullydiscrete} (right),
which are mapped with $\mathcal{T}_n : \hat I \to I_n$,
$t = \tau_n \cdot \hat t + t_m$, from $\hat I$ to $I_n$.}
\label{fig:3:basisfunctions}
\end{figure}

The primal and dual problem are discretised with appropriate
space-time variational methods, i.e. here
a piecewise constant discontinuous Galerkin method in time for the primal problem
and
a piecewise linear continuous Petrov-Galerkin method in time for the dual problem
(Fig. \ref{fig:2:fullydiscrete}),
combined with a piecewise polynomial continuous Ritz-Galerkin (FEM) method in space;
cf. for details \cite{Koecher2015} and \cite{Bruchhaeuser2018,Bangerth2003}.
The space-time cylinder $\mathcal{Q}=\Omega \times I$
is divided into non-overlapping space-time slabs
$\mathcal{Q}_n^\ell = \Omega_h^\ell \times I_n^\ell$,
with the partition of $\bar{I}=[t_0,T]$ as
$t_0 =: t_0^\ell < \cdots < t_{N^\ell}^\ell := T$,
$I_n^\ell:=(t_{m}^\ell,t_n^\ell)$, $t_{m}^\ell:=t_{n-1}^\ell$, $n=1,\dots,N^\ell$,
for the $\ell$-th loop.
On each $\mathcal{Q}_n^\ell$ consider a not necessarily conforming
partition $\mathcal{T}_{h,n}^\ell$ of $\Omega_h^\ell$
into non-overlapping elements $K_n^\ell$, denoted as
geometrical quadrilaterals for $d=2$ or hexahedrons for $d=3$.
The fully discrete primal and dual solutions are represented by
\begin{equation}
\label{eq:7:primal:fullydiscrete}
u_{\tau,h}^\ell(\boldsymbol x, t) =
\sum_{n=1}^{N^\ell} \sum_{\iota=0}^0 \sum_{j=1}^{N_\text{DoF}^{\text{primal},n,\ell}}
u^{n,\ell}_{j,\iota} \cdot
\phi_j^{\text{primal},n,\ell}(\boldsymbol x) \cdot
\zeta_{\iota}^{\text{primal},n,\ell}(t)
\end{equation}
%
\begin{equation}
\label{eq:8:dual:fullydiscrete}
\text{and}\quad
z_{\tau,h}^\ell(\boldsymbol x, t) =
\sum_{n=1}^{N^\ell} \sum_{\iota=0}^1 \sum_{j=1}^{N_\text{DoF}^{\text{dual},n,\ell}}
z^{n,\ell}_{j,\iota} \cdot
\phi_j^{\text{dual},n,\ell}(\boldsymbol x) \cdot
\xi_{\iota}^{\text{dual},n,\ell}(t)\,.
\end{equation}
In \eqref{eq:7:primal:fullydiscrete}-\eqref{eq:8:dual:fullydiscrete},
$u^{n,\ell}_{j,\iota}$ and $z^{n,\ell}_{j,\iota}$ denote
the space-time degrees of freedom (DoF) of the primal and dual problem,
$\phi_j^{\text{primal},n,\ell}$ and $\phi_j^{\text{dual},n,\ell}$
denote the global spatial trial basis functions and
$\zeta_{\iota}^{\text{primal},n,\ell}$ and $\xi_{\iota}^{\text{dual},n,\ell}$
denote the global temporal trial basis functions (Fig. \ref{fig:3:basisfunctions}).
The basis functions are defined on a reference space-time slab
$\hat{\mathcal{Q}} = (0,1)^d \times (0,1)$,
e.g. as piecewise Lagrange polynomials,
and mapped appropriately to $\mathcal{T}_{h,n}^\ell \times I_n^\ell$;
cf. \cite{Koecher2015} for details.

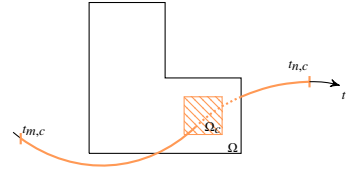
\begin{figure}[t]
\centering

\vskip1.5ex
\begin{tikzpicture}
\draw (0,0) -- (2,0) -- (2,1) -- (1,1) -- (1,2) -- (0,2) -- (0,0);

\draw[PrimalOrange]
  (1.25,.25) -- (1.75,.25) -- (1.75,.75) -- (1.25,.75) -- (1.25,.25);
\fill[pattern=north west lines, pattern color=PrimalOrange]
  (1.25,.25) -- (1.75,.25) -- (1.75,.75) -- (1.25,.75) -- (1.25,.25);

\node[anchor=south west] at (1.7,-.1) {\tiny $\Omega$};
\node[anchor=south west] at (1.4,.15) {\tiny $\Omega_c$};

\path [black,bend right=1] (-1.,.28) edge (-.9,.2);

\draw[PrimalOrange,thick] (-.9,.05+.07) -- (-.9,.20+.07);
\node[anchor=south west] at (-1,.1) {\tiny $t_{m,c}$};

\path [PrimalOrange,bend right=40,thick] (-.9,.2) edge (1.4,.35);

\path [PrimalOrange,bend left=10, densely dotted,thick] (1.4,.35) edge (2,.75);
\path [PrimalOrange,bend left=10,thick] (2,.75) edge (2.9,.95);

\draw[PrimalOrange,thick] (2.9,.05+.825) -- (2.9,.20+.825);
\node[anchor=south west] at (2.5,.95) {\tiny $t_{n,c}$};

\path [->,>=stealth',black,bend left=8] (2.9,.95) edge (3.3,.90);
\node[anchor=south west] at (3.2,.6) {\tiny $t$};

\end{tikzpicture}

\caption{Illustration of the implemented
local ($d$+$1$)-dimensional space-time control volume
$\mathcal{Q}_c(t) =
\Omega_c(t) \times I_c \subseteq \mathcal{Q}=\Omega \times I$,
with $I_c =(t_{m,c},t_{n,c})$ and $t_0 \le t_{m,c} < t_{n,c} \le T$,
for Eq. \eqref{eq:9:Jphi} and Sec. \ref{sec:3:examples}.}
\label{fig:4:LocalControlVolume}
\end{figure}

The software implements the goal functional $J : V \to \mathbb{R}$,
\begin{equation}
\label{eq:9:Jphi}
J(\psi) := \displaystyle\frac{1}{\| u - u_{\tau,h}^\ell \|_{\mathcal{Q}_c}}\,
  \displaystyle \int_{I_c} \int_{\Omega_c} \psi \cdot (u - u_{\tau,h}^\ell)\,
  \operatorname{d} \boldsymbol x\, \operatorname{d} t\,,
\end{equation}
for all $\psi \in V$,
with $u_{\tau,h}^\ell$ denoting the fully discrete primal solution of the
$\ell$-th DWR loop and
$u$ denoting the analytic solution (only possible for academic test problems)
or an approximation of the exact primal solution generated on a finer space-time
mesh or with higher-order approaches,
which aims to reach
\begin{equation}
\label{eq:10:goalTol}
\| u - u_{\tau,h}^\ell \|_{\mathcal{Q}_c} < \text{\texttt{tol}}\,,
\end{equation}
for an absolute or relative tolerance \texttt{tol} criterion,
on the space-time control volume
$\mathcal{Q}_c = \Omega_c \times I_c \subseteq \mathcal{Q}$;
cf. Fig. \ref{fig:4:LocalControlVolume}.

\textbf{Algorithm (dwr-diffusion)}.
Loop $\ell=1,\dots$ until the goal \eqref{eq:10:goalTol} is reached;
cf. Fig. \ref{fig:1:space-time-dwr}.
\textit{Solve the primal problem:} Find the coefficient vector
$\boldsymbol u^{n,\ell}_{0} = ( u^{n,\ell}_{j,0} )_j$,
$j=1, \dots, N_\text{DoF}^{\text{primal},n,\ell}$, from
\begin{equation}
\label{eq:11:primal}
(\boldsymbol M_n + \tau_n\, \boldsymbol A_n)\, \boldsymbol u^{n,\ell}_{0} =
\tau_n\, ( \boldsymbol f^{n}_{0} + \boldsymbol h^{n}_{0} )
+ \boldsymbol M_n\, (I_h\, u_{\tau,h}^\ell(\boldsymbol x, t_{n-1}))_j\,,
\end{equation}
for $n=1,\dots,N^\ell$, by marching forwardly in time through the slabs.
The mass and stiffness matrix of the primal problem on the slab $n$ are denoted
as $\boldsymbol M_n$ and $\boldsymbol A_n$, respectively,
the right hand side assemblies $\boldsymbol f^{n}_{0}$ and $\boldsymbol h^{n}_{0}$
correspond to volume forcing and inhomogeneous Neumann boundary terms and
the vector $(I_h\, u_{\tau,h}^\ell(\boldsymbol x, t_{n-1}))_j$
is the interpolation of the initial value function $u_0$ from \eqref{eq:3:model}
for $n=1$ or
the fully discrete primal solution of the previous ($n$-$1$) slab for $n>1$
on the primal finite element space of the current slab $\mathcal{Q}_n^\ell$.
Each system is modified such that the Dirichlet boundary conditions from
\eqref{eq:3:model} are applied strongly.
To set up the dual problem with the goal \eqref{eq:9:Jphi},
\textit{compute the contribution of the value} $\| u - u_{\tau,h} \|_{\mathcal{Q}_c}$
\eqref{eq:10:goalTol} with a post-processing on each slab $\mathcal{Q}_n^\ell$.
\textit{Solve the dual problem:} Find the coefficient vector
$\boldsymbol z^{n,\ell}_{0} = ( z^{n,\ell}_{j,0} )_j$,
$j=1, \dots, N_\text{DoF}^{\text{dual},n,\ell}$, from
\begin{equation}
\label{eq:12:dual}
(2\, \boldsymbol M_n^d + \tau_n\, \boldsymbol A_n^d)\, \boldsymbol z^{n,\ell}_{0} =
\tau_n\, \boldsymbol J^{n,\ell}_{0}
+ 2\, \boldsymbol M_n^d\, (I_h^d\, z_{\tau,h}^\ell(\boldsymbol x, t_{n}))_j\,,
\end{equation}
for $n=N^\ell, \dots, 1$, by marching backwardly in time through the slabs.
The mass and stiffness matrix of the dual problem on the slab $n$ are
denoted as $\boldsymbol M_n^d$ and $\boldsymbol A_n^d$, respectively,
the right hand side assembly vector $\boldsymbol J^{n,\ell}_{0}$
corresponds to the goal defined by \eqref{eq:9:Jphi}
and the vector $(I_h^d\, z_{\tau,h}^\ell(\boldsymbol x, t_{n}))_j$
is the interpolation of
the homogeneous initial value function $z_T=0$ \eqref{eq:6:dual}
for the chosen goal \eqref{eq:9:Jphi} for $n=N^\ell$ or
the fully discrete dual solution of the next ($n$+$1$) slab for $n<N^\ell$
on the dual finite element space of the current slab $\mathcal{Q}_n^\ell$.
The same type of boundary colorisation (either Dirichlet or Neumann type) is
used for the dual problem but with homogeneous boundary value functions,
even in the case of inhomogeneous primal boundary conditions.
Each system is modified such that homogeneous Dirichlet boundary conditions on
$\Gamma_D$ are applied strongly to the dual solution $z_{\tau,h}^\ell$
\eqref{eq:8:dual:fullydiscrete}.
The numerically approximated a posteriori space-time error estimate
\begin{equation}
\label{eq:13:estimator}
\tilde \eta_{\tau,h}^\ell =
\sum_{n=1}^{N^\ell} \tilde \eta^{n,\ell}\,,\quad \text{with}\quad
\tilde \eta^{n,\ell} =
\sum_{K \in \mathcal{T}_{h,n}^\ell} | \tilde \eta_K^{n,\ell} |\,,
\end{equation}
is derived from the primal and dual solutions
$u_{\tau,h}^\ell$ and $z_{\tau,h}^\ell$, respectively,
and each localised $\tilde \eta_K^{n,\ell}$, $K \in \mathcal{T}_{h,n}^\ell$,
$n=1,\dots,N^\ell$, is stored independently;
cf. for details \cite{Bruchhaeuser2018a}.
The space-time mesh refinement update is implemented as follows:
mark a space-time slab for refinement in time for which the corresponding value of
$\tilde \eta^{n,\ell}$ \eqref{eq:13:estimator}
belongs to the top fraction $0 \le \theta_\tau \le 1$ of largest values,
then, on each slab $\mathcal{Q}_n^\ell$, mark a mesh cell $K \in \mathcal{T}_{h,n}^\ell$
for $d$-dimensional isotropic refinement in space for which the corresponding value of
$| \tilde \eta_K^{n,\ell} |$ \eqref{eq:13:estimator}
belongs to the top fraction $\theta_{h,1}$ or $\theta_{h,2}$,
for a slab that is not or is marked for time refinement,
of largest values, with $0 \le \theta_{h,2} \le \theta_{h,1} \le 1$,
then execute the spatial refinement and
finally execute the temporal refinement.

\subsection{Contribution of the Software to Scientific Discovery}
\label{sec:1:3:contribution}

Modules of the \texttt{DTM++.Project},
including higher-order in time discretisations
with sophisticated solver technologies
and distributed-memory parallel implementations,
already have contributed to the process of scientific discovery for
acoustic, elastic and coupled wave propagation (\texttt{xwave}),
mass conservative transport (\texttt{meat}) and
coupled deformation with transport (\texttt{biot}) for instance;
cf. \cite{Koecher2015,Bause2017} and references therein.
The stationary predecessor of \texttt{dwr-diffusion} is used in
\cite{Bruchhaeuser2018} and the successor is used in \cite{Bruchhaeuser2018a}
for stabilised instationary convection-dominated transport problems.
The contributed software shares the modularity and flexibility of all
\texttt{DTM++.Project} modules,
it provides a freely available open-source framework for efficiently solving
problems with goal-oriented adaptivity and
will support further scientific discovery with numerical simulations
of challenging problems with physical relevance.

\subsection{Basic Setup and Usage of the Software}
\label{sec:1:4:setup}

The user needs to prepare a \texttt{deal.II} v9.0 toolchain to compile and
run the software, this can be done on any major platform by using \texttt{candi};
open a terminal and type in\\[0ex]

\begin{minipage}{.8\linewidth}
\footnotesize

\verb
git clone https://github.com/dealii/candi

\verb
cd candi && ./candi.sh
\end{minipage}\\[0ex]

\noindent
and follow the instructions. Then download, compile and run\\[0ex]

\begin{minipage}{.8\linewidth}
\footnotesize

\verb
git clone https://github.org/dtm-project/dwr-diffusion

\verb
cd dwr-diffusion && cmake . && make release

\verb
./dwr-diffusion ./input/KoecherBruchhaeuser2d.prm
\end{minipage}\\[0ex]

\noindent
with the experimental setting given in Sec. \ref{sec:3:examples}.

\subsection{Related work}
\label{sec:1:5:related}

The origin of the code is related to the \texttt{step-14} tutorial code of
\texttt{deal.II} \cite{dealIIReferenceV90} for the Laplace equation and
other \texttt{DTM++.Project} modules \cite{Koecher2015}.
Additionally, the code gallery of \cite{dealIIReferenceV90} provides some
implementations for stationary elastoplasticity problems using DWR-based
goal-oriented mesh adaptivity and
for the instationary convection-diffusion-reaction equation without adaptivity
for instance.
All of the listed related implementations can gain advantages from our work
since those are mostly specific implementations
with partly lacking documentation and, more importantly,
do not document their data structures and algorithms in an application-free way
for allowing the reuse in other frameworks straight forwardly.

\section{Software description}
\label{sec:2:software}

The software \texttt{DTM++.Project}/\texttt{dwr-diffusion}
implements the introduced algorithm of Sec. \ref{sec:1:2:ExactSciProblemSoftware}
in an object-oriented way and is shipped with an extensive in-source documentation.
\texttt{DTM++} modules are written in the \texttt{C++.17} language and
make use of the new language features since \texttt{C++.11} such as
reference counted dynamic memory allocation, range-based loops,
the \texttt{auto} specifier, strongly typed \texttt{enum} classes and others.

\begin{figure}[t]
\centering
\includegraphics[width=.65\linewidth]{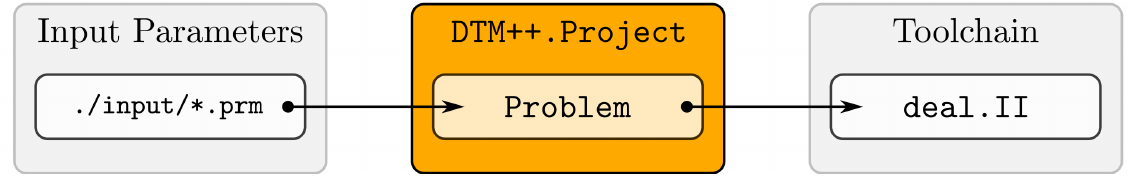}

\vskip-3.06ex
~

\caption{Pictorial \texttt{DTM++.Project} software module architecture overview.}
\label{fig:5:dtm:overview}
\end{figure}

\begin{figure}[t]
\centering

\includegraphics[width=.65\linewidth]{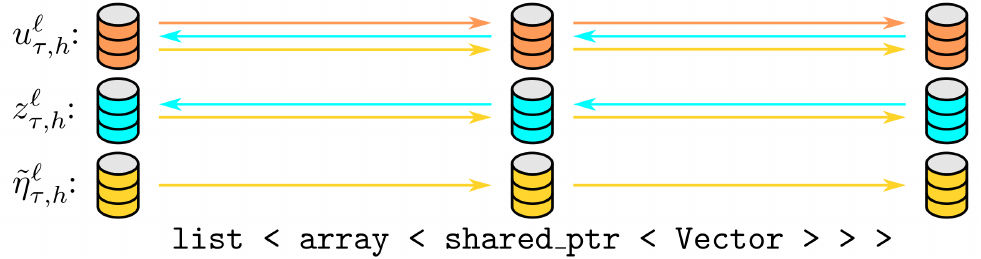}

\vskip-3.06ex
~

\caption{Pictorial overview of the \texttt{DTM++.Project} data storage management.
A list entry corresponds to a slab and stores a small array of shared pointer references.
The large data \texttt{Vector} is stored independently and can be serialised.}
\label{fig:6:dtm:dwr:storage}
\end{figure}

\begin{figure}[t]
\centering

\includegraphics[width=.7\linewidth]{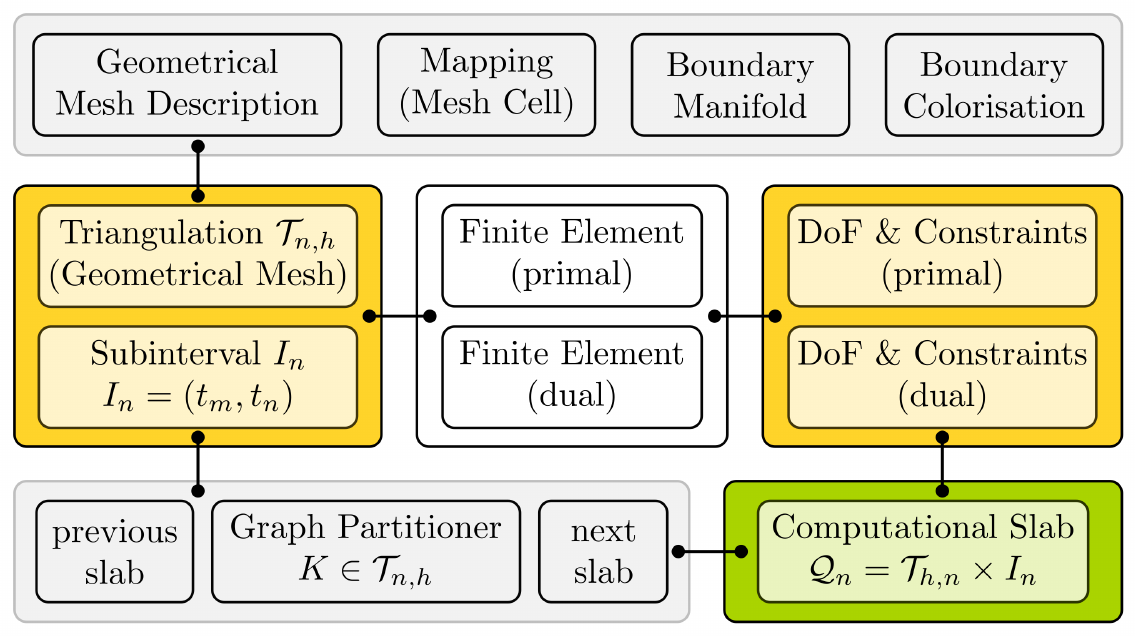}
\caption{Pictorial overview of the elements to handle a computational slab
$\mathcal{Q}_n = \mathcal{T}_{h,n} \times I_n$ implemented by the
\texttt{DTM++::Grid\_DWR} class.
Remark that loop based slab data structures $\mathcal{Q}_n^\ell$
is not stored for efficiency reasons.}
\label{fig:7:dtm:grid:spacetime}
\end{figure}

\subsection{Software Architecture}
\label{sec:2:1:software:architecture}

The general workflow to use the \texttt{dwr-diffusion} solver module is
illustrated by Fig. \ref{fig:5:dtm:overview}, it highly rely on user-driven
inputs by simple text-based parameter input files for a wide range of
similar numerical examples.
To keep the code clean for the intended user group,
the implementation of the presented algorithm from
Sec. \ref{sec:1:2:ExactSciProblemSoftware}
is done with a procedural-structured single central solver class template
\texttt{Diffusion\_DWR<dim>} : \texttt{DTM::Problem}
for slightly different time discretisations of the primal and dual problem.
All assemblers, such as for the matrices, right-hand side vectors and
even the space-time error estimate $\tilde \eta$,
are using the \texttt{deal.II} workstream technology for thread-parallel
and further \texttt{MPI+X}-parallel simulations.
Thereby, the useful and efficient auxiliary classes of the \texttt{DTM++}
suite are provided independently, such that the input parameter handling
and data output handling must not be included into the solver class.
An overview on the general \texttt{DTM++} software framework
for PDE solver frameworks can be found in \cite{Koecher2015}
and by the in-source \texttt{doxygen} documentation of the code.

\subsection{Software Key Technologies}
\label{sec:2:2:software:functionalities}

One of the key technologies of this work is the efficient handling of the
space-time DoF data storage management as given in Fig. \ref{fig:6:dtm:dwr:storage}
which allow to efficiently iterate forwardly and backwardly to compute the primal
and dual solutions as well as the error estimate.
Another important key technology is a \texttt{list} approach of the new
\texttt{DTM++::Grid\_DWR} class to handle ($d$+$1$)-dimensional space-time
computational slabs and grids as given in Fig. \ref{fig:7:dtm:grid:spacetime}
and allows for inexpensive local time refinements.

\section{Illustrative Examples}
\label{sec:3:examples}

To demonstrate the functionalities, an analytic solution $u$,
which mimics a rotating cone with a time-dependent height,
\begin{equation}
\label{eq:14:analyticSolution}
\begin{array}{l@{\,}c@{\,}l}
u(\boldsymbol x, t) &:=&
u_1 \cdot u_2\,,\,\,
\boldsymbol x = [x_1, x_2] \in \mathbb{R}^2 \text{ and }
t \in \mathbb{R}\,,\\[.5ex]
u_1(\boldsymbol x, t) &:=&
  (1 + a \cdot ( (x_1 - m_1(t))^2 + (x_2 - m_2(t))^2  ) )^{-1}\,,\\[.5ex]
u_2(t) &:=& \nu_1(t) \cdot s \cdot \arctan( \nu_2(t) )\,,
\end{array}
\end{equation}
with $m_1(t) := \frac{1}{2} + \frac{1}{4} \cos(2 \pi t)$ and
$m_2(t) := \frac{1}{2} + \frac{1}{4} \sin(2 \pi t)$, and,
$\nu_1(\hat t) := -1$,
$\nu_2(\hat t) := 5 \pi \cdot (4 \hat t - 1)$,
for $\hat t \in [0, 0.5)$ and
$\nu_1(\hat t) := 1$,
$\nu_2(\hat t) := 5 \pi \cdot (4 (\hat t-0.5) - 1)$,
for $\hat t \in [0.5, 1)$, $\hat t = t - k$,
$k \in \mathbb{N}_0$, and,
scalars $a, s \in \mathbb{R}$, $a > 0$,
is approximated on a two-dimensional L-shaped domain
(Fig. \ref{fig:4:LocalControlVolume})
for $I=(0,1.25)$ with the minimisation goal from
\eqref{eq:9:Jphi} on a time-dependent space-time control volume
$\mathcal{Q}_{c,\text{KB2d}} = \Omega_c(t) \times I_c$, with
$\Omega_c(t) =
\{ {\boldsymbol x \in \mathbb{R}^2} \,|\,
\boldsymbol x - [0.5, 0.5]^T = \hat{\boldsymbol x} + \boldsymbol m(t;r_1)\,,\,
\hat{\boldsymbol x} \in \hat \Omega\,,\,
\boldsymbol m(t; r_1) = r_1 \cdot [\cos(t),\, \sin(t)]^T \}$,
$\hat \Omega =
(\hat p^1_1,\, \hat p^2_1) \times (\hat p^1_2,\, \hat p^2_2) \subset \mathbb{R}^2$
and $I_c=(0.25, 1)$.

\begin{figure}[t]
\centering

\vskip-2ex
\includegraphics[width=.3\linewidth]{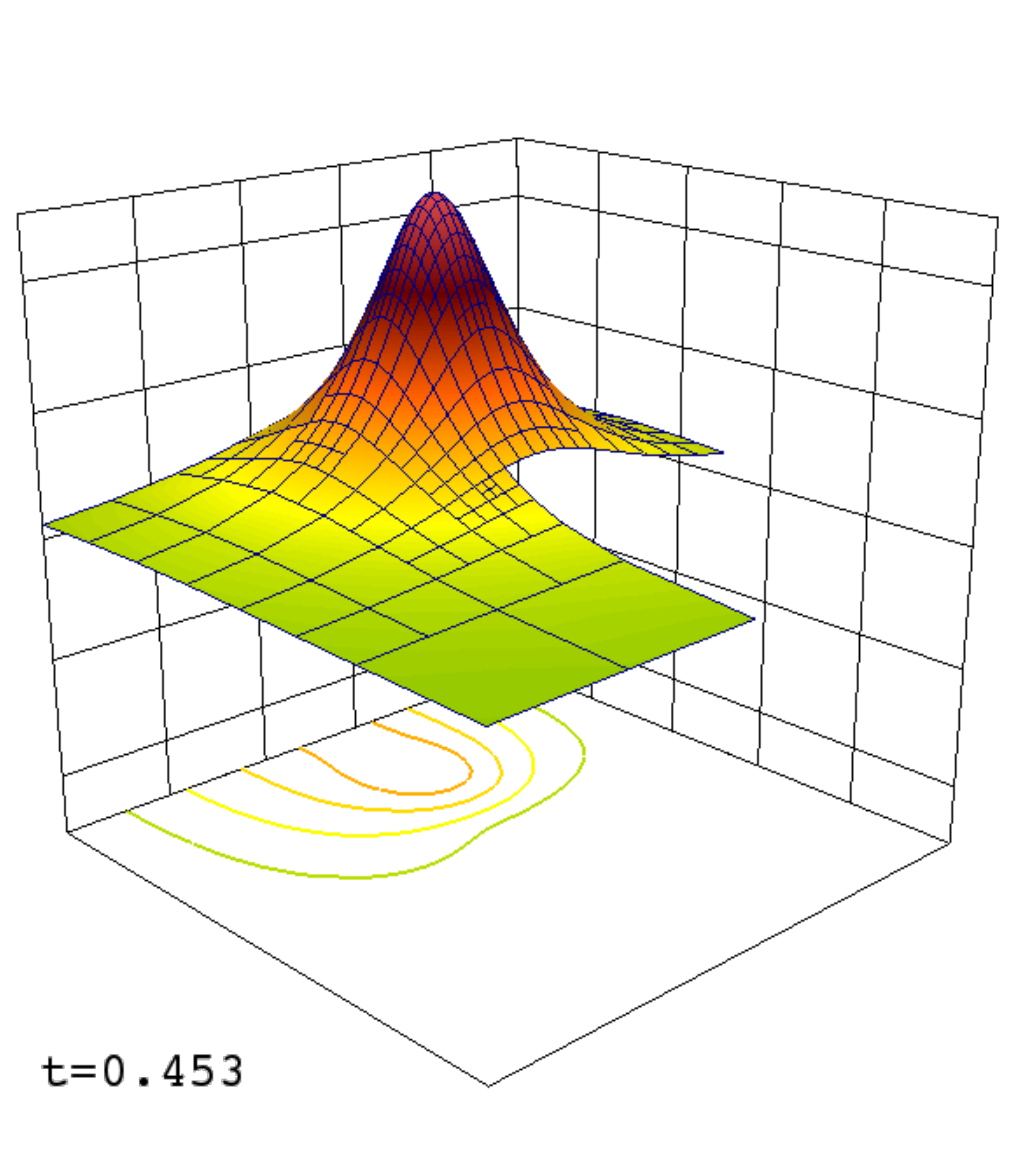}
\includegraphics[width=.3\linewidth]{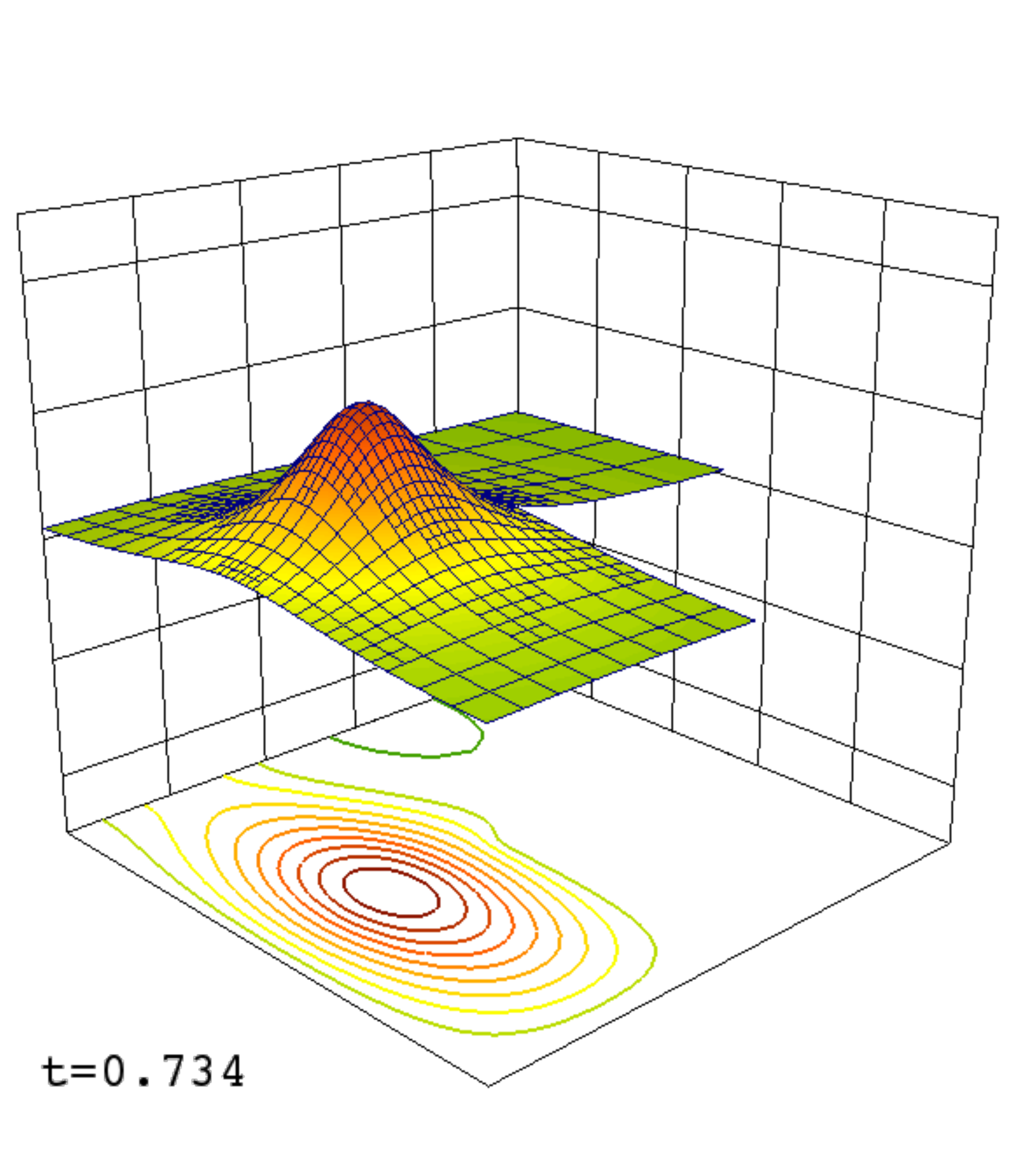}
\includegraphics[width=.3\linewidth]{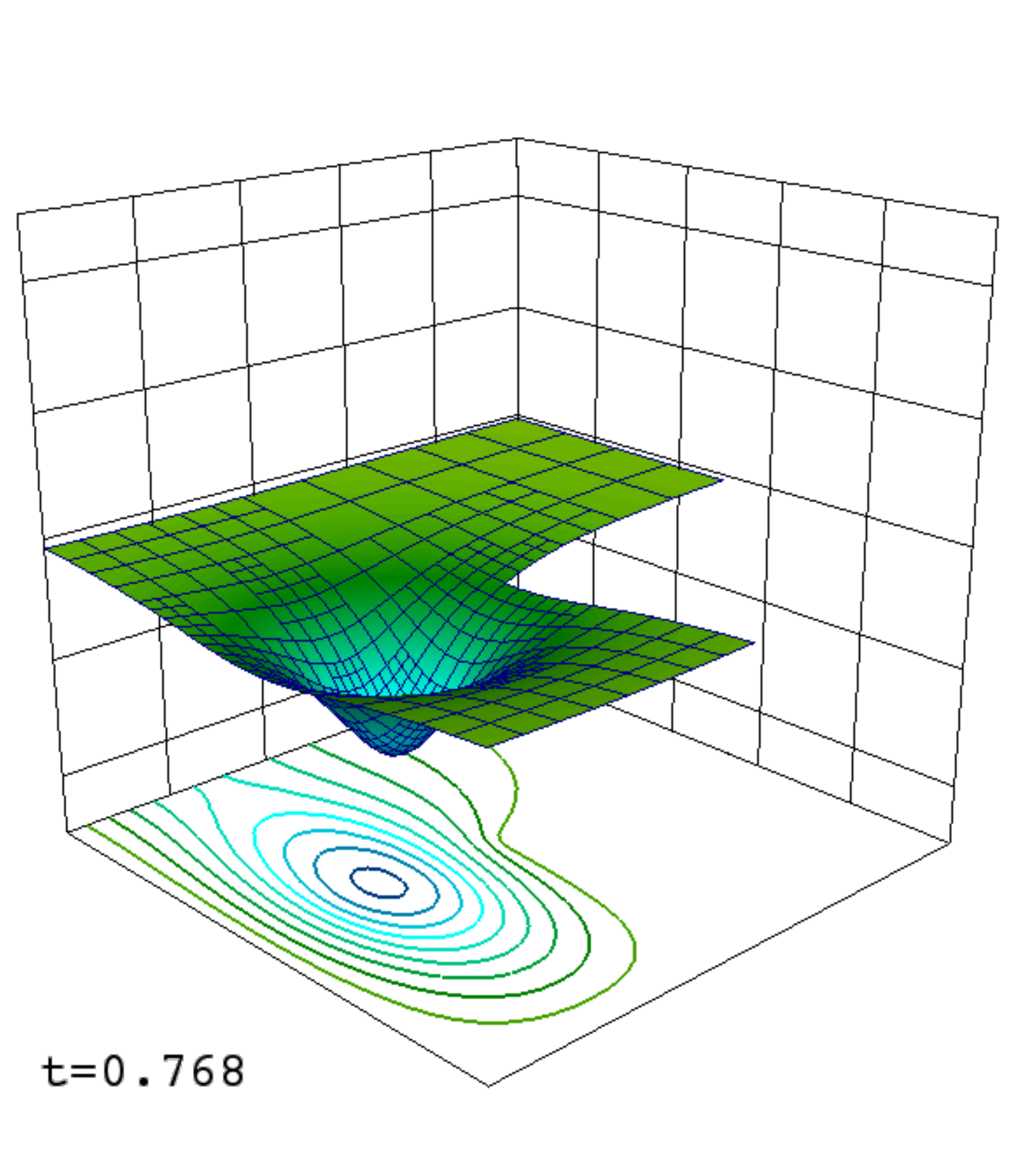}

\vskip-5ex
~

\caption{Solution profiles $u_{\tau,h}^{18}$ and
dual solution $z_{\tau,h}^{18}$ contour lines
for Sec. \ref{sec:3:examples}.}
\label{fig:8:solution}
\end{figure}

\begin{figure}[t]
\centering

\hspace*{-.11in}
\begin{tikzpicture}
\begin{footnotesize}
\begin{axis}[%
width=2.85in,
height=0.5in,
scale only axis,
ylabel={$\tau_n(I_n^{17})$},
xmin=-0.05,
xmax=1.30,
xtick={0.00,0.25,0.50,0.75,1.00,1.25},
xticklabels={0.00,0.25,0.50,0.75,1.00},
ymode=log,
ymin=7e-04,
ymax=4e-01,
yminorticks=true,
]

\addplot[
color=black,
solid,
line width=0.3pt
]
table[row sep=crcr]{
0.00 0.0001 \\
0.00 0.5000 \\
};

\addplot[
color=black,
solid,
line width=0.3pt
]
table[row sep=crcr]{
1.25 0.0001 \\
1.25 0.5000 \\
};

\addplot[
color=black,
densely dotted,
line width=0.5pt
]
table[row sep=crcr]{
0.25 0.0001 \\
0.25 0.5000 \\
};

\addplot[
color=black,
densely dotted,
line width=0.5pt
]
table[row sep=crcr]{
1.00 0.0001 \\
1.00 0.5000 \\
};


\addplot[
color=SlabGreen,
line width=0.5pt
]
table[row sep=crcr]{
0.0000 0.0625 \\
0.0625 0.0625 \\
};

\addplot[
color=SlabGreen,
line width=0.5pt
]
table[row sep=crcr]{
0.0625  0.03125 \\
0.125   0.03125 \\
};

\addplot[
color=SlabGreen,
line width=0.5pt
]
table[row sep=crcr]{
0.125    0.015625 \\
0.203125 0.015625 \\
};

\addplot[
color=SlabGreen,
line width=0.5pt
]
table[row sep=crcr]{
0.203125 0.0078125 \\
0.25     0.0078125 \\
};

\addplot[
color=SlabGreen,
line width=0.5pt
]
table[row sep=crcr]{
0.250977 0.000976562 \\
0.251953 0.000976562 \\
};

\addplot[
color=SlabGreen,
line width=0.5pt
]
table[row sep=crcr]{
0.251953 0.00195312 \\
0.257812 0.00195312 \\
};

\addplot[
color=SlabGreen,
line width=0.5pt
]
table[row sep=crcr]{
0.257812 0.00390625 \\
0.273438 0.00390625 \\
};

\addplot[
color=SlabGreen,
line width=0.5pt
]
table[row sep=crcr]{
0.273438 0.0078125 \\
0.28125  0.0078125 \\
};

\addplot[
color=SlabGreen,
line width=0.5pt
]
table[row sep=crcr]{
0.28125  0.00390625 \\
0.515625 0.00390625 \\
};

\addplot[
color=SlabGreen,
line width=0.5pt
]
table[row sep=crcr]{
0.515625 0.00195312 \\
0.558594 0.00195312 \\
};

\addplot[
color=SlabGreen,
line width=0.5pt
]
table[row sep=crcr]{
0.558594 0.00390625 \\
0.761719 0.00390625 \\
};

\addplot[
color=SlabGreen,
line width=0.5pt
]
table[row sep=crcr]{
0.761719 0.00195312 \\
0.765625 0.00195312 \\
};

\addplot[
color=SlabGreen,
line width=0.5pt
]
table[row sep=crcr]{
0.765625 0.00390625 \\
0.769531 0.00390625 \\
};

\addplot[
color=SlabGreen,
line width=0.5pt
]
table[row sep=crcr]{
0.769531 0.00195312 \\
0.78125  0.00195312 \\
};

\addplot[
color=SlabGreen,
line width=0.5pt
]
table[row sep=crcr]{
0.78125  0.00390625 \\
0.785156 0.00390625 \\
};

\addplot[
color=SlabGreen,
line width=0.5pt
]
table[row sep=crcr]{
0.785156 0.00195312 \\
0.792969 0.00195312 \\
};

\addplot[
color=SlabGreen,
line width=0.5pt
]
table[row sep=crcr]{
0.792969 0.00390625 \\
0.9375   0.00390625 \\
};

\addplot[
color=SlabGreen,
line width=0.5pt
]
table[row sep=crcr]{
0.9375   0.00195312 \\
0.941406 0.00195312 \\
};

\addplot[
color=SlabGreen,
line width=0.5pt
]
table[row sep=crcr]{
0.941406 0.00390625 \\
0.945312 0.00390625 \\
};

\addplot[
color=SlabGreen,
line width=0.5pt
]
table[row sep=crcr]{
0.945312 0.00195312 \\
0.949219 0.00195312 \\
};

\addplot[
color=SlabGreen,
line width=0.5pt
]
table[row sep=crcr]{
0.949219 0.00390625 \\
0.992188 0.00390625 \\
};

\addplot[
color=SlabGreen,
line width=0.5pt
]
table[row sep=crcr]{
0.992188 0.0078125 \\
1        0.0078125 \\
};

\addplot[
color=SlabGreen,
line width=0.5pt
]
table[row sep=crcr]{
1    0.25 \\
1.25 0.25 \\
};


\addplot[
color=SlabGreen!80!black,
only marks,
mark=*,
mark size = 0.5,
mark options={solid}
]
table[row sep=crcr]{
0.0625 0.0625 \\
0.09375 0.03125 \\
0.125 0.03125 \\
0.140625 0.015625 \\
0.15625 0.015625 \\
0.171875 0.015625 \\
0.1875 0.015625 \\
0.203125 0.015625 \\
0.210938 0.0078125 \\
0.21875 0.0078125 \\
0.226562 0.0078125 \\
0.234375 0.0078125 \\
0.242188 0.0078125 \\
0.25 0.0078125 \\
0.250977 0.000976562 \\
0.251953 0.000976562 \\
0.253906 0.00195312 \\
0.255859 0.00195312 \\
0.257812 0.00195312 \\
0.261719 0.00390625 \\
0.265625 0.00390625 \\
0.269531 0.00390625 \\
0.273438 0.00390625 \\
0.28125 0.0078125 \\
0.285156 0.00390625 \\
0.289062 0.00390625 \\
0.292969 0.00390625 \\
0.296875 0.00390625 \\
0.300781 0.00390625 \\
0.304688 0.00390625 \\
0.308594 0.00390625 \\
0.3125 0.00390625 \\
0.316406 0.00390625 \\
0.320312 0.00390625 \\
0.324219 0.00390625 \\
0.328125 0.00390625 \\
0.332031 0.00390625 \\
0.335938 0.00390625 \\
0.339844 0.00390625 \\
0.34375 0.00390625 \\
0.347656 0.00390625 \\
0.351562 0.00390625 \\
0.355469 0.00390625 \\
0.359375 0.00390625 \\
0.363281 0.00390625 \\
0.367188 0.00390625 \\
0.371094 0.00390625 \\
0.375 0.00390625 \\
0.378906 0.00390625 \\
0.382812 0.00390625 \\
0.386719 0.00390625 \\
0.390625 0.00390625 \\
0.394531 0.00390625 \\
0.398438 0.00390625 \\
0.402344 0.00390625 \\
0.40625 0.00390625 \\
0.410156 0.00390625 \\
0.414062 0.00390625 \\
0.417969 0.00390625 \\
0.421875 0.00390625 \\
0.425781 0.00390625 \\
0.429688 0.00390625 \\
0.433594 0.00390625 \\
0.4375 0.00390625 \\
0.441406 0.00390625 \\
0.445312 0.00390625 \\
0.449219 0.00390625 \\
0.453125 0.00390625 \\
0.457031 0.00390625 \\
0.460938 0.00390625 \\
0.464844 0.00390625 \\
0.46875 0.00390625 \\
0.472656 0.00390625 \\
0.476562 0.00390625 \\
0.480469 0.00390625 \\
0.484375 0.00390625 \\
0.488281 0.00390625 \\
0.492188 0.00390625 \\
0.496094 0.00390625 \\
0.5 0.00390625 \\
0.503906 0.00390625 \\
0.507812 0.00390625 \\
0.511719 0.00390625 \\
0.515625 0.00390625 \\
0.517578 0.00195312 \\
0.519531 0.00195312 \\
0.521484 0.00195312 \\
0.523438 0.00195312 \\
0.525391 0.00195312 \\
0.527344 0.00195312 \\
0.529297 0.00195312 \\
0.53125 0.00195312 \\
0.533203 0.00195312 \\
0.535156 0.00195312 \\
0.537109 0.00195312 \\
0.539062 0.00195312 \\
0.541016 0.00195312 \\
0.542969 0.00195312 \\
0.544922 0.00195312 \\
0.546875 0.00195312 \\
0.548828 0.00195312 \\
0.550781 0.00195312 \\
0.552734 0.00195312 \\
0.554688 0.00195312 \\
0.556641 0.00195312 \\
0.558594 0.00195312 \\
0.5625 0.00390625 \\
0.566406 0.00390625 \\
0.570312 0.00390625 \\
0.574219 0.00390625 \\
0.578125 0.00390625 \\
0.582031 0.00390625 \\
0.585938 0.00390625 \\
0.589844 0.00390625 \\
0.59375 0.00390625 \\
0.597656 0.00390625 \\
0.601562 0.00390625 \\
0.605469 0.00390625 \\
0.609375 0.00390625 \\
0.613281 0.00390625 \\
0.617188 0.00390625 \\
0.621094 0.00390625 \\
0.625 0.00390625 \\
0.628906 0.00390625 \\
0.632812 0.00390625 \\
0.636719 0.00390625 \\
0.640625 0.00390625 \\
0.644531 0.00390625 \\
0.648438 0.00390625 \\
0.652344 0.00390625 \\
0.65625 0.00390625 \\
0.660156 0.00390625 \\
0.664062 0.00390625 \\
0.667969 0.00390625 \\
0.671875 0.00390625 \\
0.675781 0.00390625 \\
0.679688 0.00390625 \\
0.683594 0.00390625 \\
0.6875 0.00390625 \\
0.691406 0.00390625 \\
0.695312 0.00390625 \\
0.699219 0.00390625 \\
0.703125 0.00390625 \\
0.707031 0.00390625 \\
0.710938 0.00390625 \\
0.714844 0.00390625 \\
0.71875 0.00390625 \\
0.722656 0.00390625 \\
0.726562 0.00390625 \\
0.730469 0.00390625 \\
0.734375 0.00390625 \\
0.738281 0.00390625 \\
0.742188 0.00390625 \\
0.746094 0.00390625 \\
0.75 0.00390625 \\
0.753906 0.00390625 \\
0.757812 0.00390625 \\
0.761719 0.00390625 \\
0.763672 0.00195312 \\
0.765625 0.00195312 \\
0.769531 0.00390625 \\
0.771484 0.00195312 \\
0.773438 0.00195312 \\
0.775391 0.00195312 \\
0.777344 0.00195312 \\
0.779297 0.00195312 \\
0.78125 0.00195312 \\
0.785156 0.00390625 \\
0.787109 0.00195312 \\
0.789062 0.00195312 \\
0.791016 0.00195312 \\
0.792969 0.00195312 \\
0.796875 0.00390625 \\
0.800781 0.00390625 \\
0.804688 0.00390625 \\
0.808594 0.00390625 \\
0.8125 0.00390625 \\
0.816406 0.00390625 \\
0.820312 0.00390625 \\
0.824219 0.00390625 \\
0.828125 0.00390625 \\
0.832031 0.00390625 \\
0.835938 0.00390625 \\
0.839844 0.00390625 \\
0.84375 0.00390625 \\
0.847656 0.00390625 \\
0.851562 0.00390625 \\
0.855469 0.00390625 \\
0.859375 0.00390625 \\
0.863281 0.00390625 \\
0.867188 0.00390625 \\
0.871094 0.00390625 \\
0.875 0.00390625 \\
0.878906 0.00390625 \\
0.882812 0.00390625 \\
0.886719 0.00390625 \\
0.890625 0.00390625 \\
0.894531 0.00390625 \\
0.898438 0.00390625 \\
0.902344 0.00390625 \\
0.90625 0.00390625 \\
0.910156 0.00390625 \\
0.914062 0.00390625 \\
0.917969 0.00390625 \\
0.921875 0.00390625 \\
0.925781 0.00390625 \\
0.929688 0.00390625 \\
0.933594 0.00390625 \\
0.9375 0.00390625 \\
0.939453 0.00195312 \\
0.941406 0.00195312 \\
0.945312 0.00390625 \\
0.947266 0.00195312 \\
0.949219 0.00195312 \\
0.953125 0.00390625 \\
0.957031 0.00390625 \\
0.960938 0.00390625 \\
0.964844 0.00390625 \\
0.96875 0.00390625 \\
0.972656 0.00390625 \\
0.976562 0.00390625 \\
0.980469 0.00390625 \\
0.984375 0.00390625 \\
0.988281 0.00390625 \\
0.992188 0.00390625 \\
1 0.0078125 \\
1.25 0.25 \\
};

\end{axis}

\node at (2.8in,-0.08in) {$t$};
\end{footnotesize}
\end{tikzpicture}%

\hspace*{-.11in}
\begin{tikzpicture}
\begin{footnotesize}
\begin{axis}[%
width=2.85in,
height=0.5in,
scale only axis,
ylabel={$\tau_n(I_n^{18})$},
xmin=-0.05,
xmax=1.30,
xtick={0.00,0.25,0.50,0.75,1.00,1.25},
xticklabels={0.00,0.25,0.50,0.75,1.00},
ymode=log,
ymin=7e-04,
ymax=4e-01,
yminorticks=true,
]

\addplot[
color=black,
solid,
line width=0.3pt
]
table[row sep=crcr]{
0.00 0.0001 \\
0.00 0.5000 \\
};

\addplot[
color=black,
solid,
line width=0.3pt
]
table[row sep=crcr]{
1.25 0.0001 \\
1.25 0.5000 \\
};

\addplot[
color=black,
densely dotted,
line width=0.5pt
]
table[row sep=crcr]{
0.25 0.0001 \\
0.25 0.5000 \\
};

\addplot[
color=black,
densely dotted,
line width=0.5pt
]
table[row sep=crcr]{
1.00 0.0001 \\
1.00 0.5000 \\
};


\addplot[
color=SlabGreen,
line width=0.5pt
]
table[row sep=crcr]{
0.0    0.0625 \\
0.0625 0.0625 \\
};

\addplot[
color=SlabGreen,
line width=0.5pt
]
table[row sep=crcr]{
0.0625  0.03125 \\
0.125   0.03125 \\
};

\addplot[
color=SlabGreen,
line width=0.5pt
]
table[row sep=crcr]{
0.125    0.015625 \\
0.203125 0.015625 \\
};

\addplot[
color=SlabGreen,
line width=0.5pt
]
table[row sep=crcr]{
0.203125 0.0078125 \\
0.234375 0.0078125 \\
};

\addplot[
color=SlabGreen,
line width=0.5pt
]
table[row sep=crcr]{
0.234375 0.00390625 \\
0.25     0.00390625 \\
};

\addplot[
color=SlabGreen,
line width=0.5pt
]
table[row sep=crcr]{
0.25     0.000976562 \\
0.251953 0.000976562 \\
};

\addplot[
color=SlabGreen,
line width=0.5pt
]
table[row sep=crcr]{
0.251953 0.00195312 \\
0.257812 0.00195312 \\
};

\addplot[
color=SlabGreen,
line width=0.5pt
]
table[row sep=crcr]{
0.257812 0.00390625 \\
0.46875  0.00390625 \\
};

\addplot[
color=SlabGreen,
line width=0.5pt
]
table[row sep=crcr]{
0.46875  0.00195312 \\
0.476562 0.00195312 \\
};

\addplot[
color=SlabGreen,
line width=0.5pt
]
table[row sep=crcr]{
0.476562 0.00390625 \\
0.484375 0.00390625 \\
};

\addplot[
color=SlabGreen,
line width=0.5pt
]
table[row sep=crcr]{
0.484375 0.00195312 \\
0.5      0.00195312 \\
};

\addplot[
color=SlabGreen,
line width=0.5pt
]
table[row sep=crcr]{
0.5      0.00390625 \\
0.515625 0.00390625 \\
};

\addplot[
color=SlabGreen,
line width=0.5pt
]
table[row sep=crcr]{
0.515625 0.00195312 \\
0.5625   0.00195312 \\
};

\addplot[
color=SlabGreen,
line width=0.5pt
]
table[row sep=crcr]{
0.5625   0.00390625 \\
0.625    0.00390625 \\
};

\addplot[
color=SlabGreen,
line width=0.5pt
]
table[row sep=crcr]{
0.625    0.00195312 \\
0.6875   0.00195312 \\
};

\addplot[
color=SlabGreen,
line width=0.5pt
]
table[row sep=crcr]{
0.6875   0.00390625 \\
0.703125 0.00390625 \\
};

\addplot[
color=SlabGreen,
line width=0.5pt
]
table[row sep=crcr]{
0.703125 0.00195312 \\
0.742188 0.00195312 \\
};

\addplot[
color=SlabGreen,
line width=0.5pt
]
table[row sep=crcr]{
0.742188 0.00390625 \\
0.75     0.00390625 \\
};

\addplot[
color=SlabGreen,
line width=0.5pt
]
table[row sep=crcr]{
0.75     0.00195312 \\
0.8125   0.00195312 \\
};

\addplot[
color=SlabGreen,
line width=0.5pt
]
table[row sep=crcr]{
0.8125   0.00390625 \\
0.820312 0.00390625 \\
};

\addplot[
color=SlabGreen,
line width=0.5pt
]
table[row sep=crcr]{
0.820312 0.00195312 \\
0.835938 0.00195312 \\
};

\addplot[
color=SlabGreen,
line width=0.5pt
]
table[row sep=crcr]{
0.835938 0.00390625 \\
0.875    0.00390625 \\
};

\addplot[
color=SlabGreen,
line width=0.5pt
]
table[row sep=crcr]{
0.875    0.00195312 \\
0.917969 0.00195312 \\
};

\addplot[
color=SlabGreen,
line width=0.5pt
]
table[row sep=crcr]{
0.917969 0.00390625 \\
0.921875 0.00390625 \\
};

\addplot[
color=SlabGreen,
line width=0.5pt
]
table[row sep=crcr]{
0.921875 0.00195312 \\
0.925781 0.00195312 \\
};

\addplot[
color=SlabGreen,
line width=0.5pt
]
table[row sep=crcr]{
0.925781 0.00390625 \\
0.9375   0.00390625 \\
};

\addplot[
color=SlabGreen,
line width=0.5pt
]
table[row sep=crcr]{
0.9375   0.00195312 \\
0.964844 0.00195312 \\
};

\addplot[
color=SlabGreen,
line width=0.5pt
]
table[row sep=crcr]{
0.964844 0.00390625 \\
1        0.00390625 \\
};

\addplot[
color=SlabGreen,
line width=0.5pt
]
table[row sep=crcr]{
1    0.25 \\
1.25 0.25 \\
};


\addplot[
color=SlabGreen!60!black,
only marks,
mark=*,
mark size = 0.5,
mark options={solid}
]
table[row sep=crcr]{
0.0625 0.0625 \\
0.09375 0.03125 \\
0.125 0.03125 \\
0.140625 0.015625 \\
0.15625 0.015625 \\
0.171875 0.015625 \\
0.1875 0.015625 \\
0.203125 0.015625 \\
0.210938 0.0078125 \\
0.21875 0.0078125 \\
0.226562 0.0078125 \\
0.234375 0.0078125 \\
0.238281 0.00390625 \\
0.242188 0.00390625 \\
0.246094 0.00390625 \\
0.25 0.00390625 \\
0.250977 0.000976562 \\
0.251953 0.000976562 \\
0.253906 0.00195312 \\
0.255859 0.00195312 \\
0.257812 0.00195312 \\
0.261719 0.00390625 \\
0.265625 0.00390625 \\
0.269531 0.00390625 \\
0.273438 0.00390625 \\
0.277344 0.00390625 \\
0.28125 0.00390625 \\
0.285156 0.00390625 \\
0.289062 0.00390625 \\
0.292969 0.00390625 \\
0.296875 0.00390625 \\
0.300781 0.00390625 \\
0.304688 0.00390625 \\
0.308594 0.00390625 \\
0.3125 0.00390625 \\
0.316406 0.00390625 \\
0.320312 0.00390625 \\
0.324219 0.00390625 \\
0.328125 0.00390625 \\
0.332031 0.00390625 \\
0.335938 0.00390625 \\
0.339844 0.00390625 \\
0.34375 0.00390625 \\
0.347656 0.00390625 \\
0.351562 0.00390625 \\
0.355469 0.00390625 \\
0.359375 0.00390625 \\
0.363281 0.00390625 \\
0.367188 0.00390625 \\
0.371094 0.00390625 \\
0.375 0.00390625 \\
0.378906 0.00390625 \\
0.382812 0.00390625 \\
0.386719 0.00390625 \\
0.390625 0.00390625 \\
0.394531 0.00390625 \\
0.398438 0.00390625 \\
0.402344 0.00390625 \\
0.40625 0.00390625 \\
0.410156 0.00390625 \\
0.414062 0.00390625 \\
0.417969 0.00390625 \\
0.421875 0.00390625 \\
0.425781 0.00390625 \\
0.429688 0.00390625 \\
0.433594 0.00390625 \\
0.4375 0.00390625 \\
0.441406 0.00390625 \\
0.445312 0.00390625 \\
0.449219 0.00390625 \\
0.453125 0.00390625 \\
0.457031 0.00390625 \\
0.460938 0.00390625 \\
0.464844 0.00390625 \\
0.46875 0.00390625 \\
0.470703 0.00195312 \\
0.472656 0.00195312 \\
0.474609 0.00195312 \\
0.476562 0.00195312 \\
0.480469 0.00390625 \\
0.484375 0.00390625 \\
0.486328 0.00195312 \\
0.488281 0.00195312 \\
0.490234 0.00195312 \\
0.492188 0.00195312 \\
0.494141 0.00195312 \\
0.496094 0.00195312 \\
0.498047 0.00195312 \\
0.5 0.00195312 \\
0.503906 0.00390625 \\
0.507812 0.00390625 \\
0.511719 0.00390625 \\
0.515625 0.00390625 \\
0.517578 0.00195312 \\
0.519531 0.00195312 \\
0.521484 0.00195312 \\
0.523438 0.00195312 \\
0.525391 0.00195312 \\
0.527344 0.00195312 \\
0.529297 0.00195312 \\
0.53125 0.00195312 \\
0.533203 0.00195312 \\
0.535156 0.00195312 \\
0.537109 0.00195312 \\
0.539062 0.00195312 \\
0.541016 0.00195312 \\
0.542969 0.00195312 \\
0.544922 0.00195312 \\
0.546875 0.00195312 \\
0.548828 0.00195312 \\
0.550781 0.00195312 \\
0.552734 0.00195312 \\
0.554688 0.00195312 \\
0.556641 0.00195312 \\
0.558594 0.00195312 \\
0.560547 0.00195312 \\
0.5625 0.00195312 \\
0.566406 0.00390625 \\
0.570312 0.00390625 \\
0.574219 0.00390625 \\
0.578125 0.00390625 \\
0.582031 0.00390625 \\
0.585938 0.00390625 \\
0.589844 0.00390625 \\
0.59375 0.00390625 \\
0.597656 0.00390625 \\
0.601562 0.00390625 \\
0.605469 0.00390625 \\
0.609375 0.00390625 \\
0.613281 0.00390625 \\
0.617188 0.00390625 \\
0.621094 0.00390625 \\
0.625 0.00390625 \\
0.626953 0.00195312 \\
0.628906 0.00195312 \\
0.630859 0.00195312 \\
0.632812 0.00195312 \\
0.634766 0.00195312 \\
0.636719 0.00195312 \\
0.638672 0.00195312 \\
0.640625 0.00195312 \\
0.642578 0.00195312 \\
0.644531 0.00195312 \\
0.646484 0.00195312 \\
0.648438 0.00195312 \\
0.650391 0.00195312 \\
0.652344 0.00195312 \\
0.654297 0.00195312 \\
0.65625 0.00195312 \\
0.658203 0.00195312 \\
0.660156 0.00195312 \\
0.662109 0.00195312 \\
0.664062 0.00195312 \\
0.666016 0.00195312 \\
0.667969 0.00195312 \\
0.669922 0.00195312 \\
0.671875 0.00195312 \\
0.673828 0.00195312 \\
0.675781 0.00195312 \\
0.677734 0.00195312 \\
0.679688 0.00195312 \\
0.681641 0.00195312 \\
0.683594 0.00195312 \\
0.685547 0.00195312 \\
0.6875 0.00195312 \\
0.691406 0.00390625 \\
0.695312 0.00390625 \\
0.699219 0.00390625 \\
0.703125 0.00390625 \\
0.705078 0.00195312 \\
0.707031 0.00195312 \\
0.708984 0.00195312 \\
0.710938 0.00195312 \\
0.712891 0.00195312 \\
0.714844 0.00195312 \\
0.716797 0.00195312 \\
0.71875 0.00195312 \\
0.720703 0.00195312 \\
0.722656 0.00195312 \\
0.724609 0.00195312 \\
0.726562 0.00195312 \\
0.728516 0.00195312 \\
0.730469 0.00195312 \\
0.732422 0.00195312 \\
0.734375 0.00195312 \\
0.736328 0.00195312 \\
0.738281 0.00195312 \\
0.740234 0.00195312 \\
0.742188 0.00195312 \\
0.746094 0.00390625 \\
0.75 0.00390625 \\
0.751953 0.00195312 \\
0.753906 0.00195312 \\
0.755859 0.00195312 \\
0.757812 0.00195312 \\
0.759766 0.00195312 \\
0.761719 0.00195312 \\
0.763672 0.00195312 \\
0.765625 0.00195312 \\
0.767578 0.00195312 \\
0.769531 0.00195312 \\
0.771484 0.00195312 \\
0.773438 0.00195312 \\
0.775391 0.00195312 \\
0.777344 0.00195312 \\
0.779297 0.00195312 \\
0.78125 0.00195312 \\
0.783203 0.00195312 \\
0.785156 0.00195312 \\
0.787109 0.00195312 \\
0.789062 0.00195312 \\
0.791016 0.00195312 \\
0.792969 0.00195312 \\
0.794922 0.00195312 \\
0.796875 0.00195312 \\
0.798828 0.00195312 \\
0.800781 0.00195312 \\
0.802734 0.00195312 \\
0.804688 0.00195312 \\
0.806641 0.00195312 \\
0.808594 0.00195312 \\
0.810547 0.00195312 \\
0.8125 0.00195312 \\
0.816406 0.00390625 \\
0.820312 0.00390625 \\
0.822266 0.00195312 \\
0.824219 0.00195312 \\
0.826172 0.00195312 \\
0.828125 0.00195312 \\
0.830078 0.00195312 \\
0.832031 0.00195312 \\
0.833984 0.00195312 \\
0.835938 0.00195312 \\
0.839844 0.00390625 \\
0.84375 0.00390625 \\
0.847656 0.00390625 \\
0.851562 0.00390625 \\
0.855469 0.00390625 \\
0.859375 0.00390625 \\
0.863281 0.00390625 \\
0.867188 0.00390625 \\
0.871094 0.00390625 \\
0.875 0.00390625 \\
0.876953 0.00195312 \\
0.878906 0.00195312 \\
0.880859 0.00195312 \\
0.882812 0.00195312 \\
0.884766 0.00195312 \\
0.886719 0.00195312 \\
0.888672 0.00195312 \\
0.890625 0.00195312 \\
0.892578 0.00195312 \\
0.894531 0.00195312 \\
0.896484 0.00195312 \\
0.898438 0.00195312 \\
0.900391 0.00195312 \\
0.902344 0.00195312 \\
0.904297 0.00195312 \\
0.90625 0.00195312 \\
0.908203 0.00195312 \\
0.910156 0.00195312 \\
0.912109 0.00195312 \\
0.914062 0.00195312 \\
0.916016 0.00195312 \\
0.917969 0.00195312 \\
0.921875 0.00390625 \\
0.923828 0.00195312 \\
0.925781 0.00195312 \\
0.929688 0.00390625 \\
0.933594 0.00390625 \\
0.9375 0.00390625 \\
0.939453 0.00195312 \\
0.941406 0.00195312 \\
0.943359 0.00195312 \\
0.945312 0.00195312 \\
0.947266 0.00195312 \\
0.949219 0.00195312 \\
0.951172 0.00195312 \\
0.953125 0.00195312 \\
0.955078 0.00195312 \\
0.957031 0.00195312 \\
0.958984 0.00195312 \\
0.960938 0.00195312 \\
0.962891 0.00195312 \\
0.964844 0.00195312 \\
0.96875 0.00390625 \\
0.972656 0.00390625 \\
0.976562 0.00390625 \\
0.980469 0.00390625 \\
0.984375 0.00390625 \\
0.988281 0.00390625 \\
0.992188 0.00390625 \\
0.996094 0.00390625 \\
1 0.00390625 \\
1.25 0.25 \\
};


\end{axis}

\node at (2.8in,-0.08in) {$t$};
\end{footnotesize}
\end{tikzpicture}

\vskip-4ex
~

\caption{Distribution of $\tau_n$ of the space-time slabs $\mathcal{Q}_n^\ell$,
$\ell \in \{17,18\}$, for Sec. \ref{sec:3:examples}.}
\label{fig:9:tauDistribution}
\end{figure}
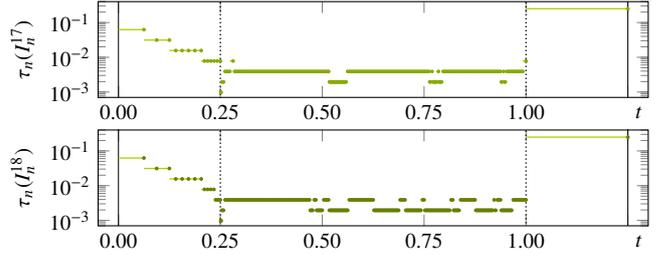

The setup uses
$\Omega_h=\{ (0,1)^2 \} \setminus \{ (0.5,1)^2 \}$,
$\Gamma_N = \{ \boldsymbol x \in \bar \Omega_h \,|\, x_1 = 0\}$,
$\Gamma_D = \partial \bar \Omega_h \setminus \Gamma_N$.
%
The functions $u_0$, $f$, $g$ and $h$ of \eqref{eq:3:model}
are derived from the analytic solution \eqref{eq:14:analyticSolution},
the coefficient functions are set to $\rho=0.8$ and $\epsilon=1.2$, and,
the scalars are set to $a=50$ and $s=-0.3333$.
The space-time triangulation for $\ell=1$ is composed of
$N^1=5$ slabs having 3 spatial cells each.
The goal tolerance is
$\text{\texttt{tol}} = 10^{-2} \cdot \| u - u_{\tau,h}^1 \|_{\mathcal{Q}_c} $.
Solution profiles for $u_{\tau,h}$ and contour lines of $z_{\tau,h}$
are illustrated in Fig. \ref{fig:8:solution}.
Selected convergence and error estimator results as well as
a quality measurement
$I_{\text{eff}} := | \tilde \eta_{\tau,h}^\ell / (J(u)-J(u_{\tau,h}^\ell)) |$,
where $I_{\text{eff}}=1$ is optimal,
are given by Tab. \ref{tab:1:convergence}.
%
%
\begin{table}[h]
\centering
\footnotesize

\begin{tabular}{rrr|c c c}
$\ell$ & $N^\ell$ &
$| \mathcal{T}_{h,\star}^\ell |$ &
$\| u - u_{\tau,h}^\ell \|_{\mathcal{Q}_c}$ &
$\tilde \eta_{\tau,h}^\ell$ &
I$_{\text{eff}}$\\
\hline
 1 &   5 &   3 & 6.070e-02 & 1.994e-02 & 0.33\\
 2 &   6 &  12 & 2.634e-02 & 2.451e-02 & 0.93\\
\multicolumn{3}{c|}{$\cdots$} & \multicolumn{3}{c}{$\cdots$} \\
17 & 227 & 441 & 7.304e-04 & 2.030e-04 & 0.28\\
18 & 295 & 603 & 5.744e-04 & 1.723e-04 & 0.30\\
\end{tabular}

\caption{Selected convergence and error estimator results of loop $\ell$,
having $N^\ell$ slabs and
the max. number of mesh cells $| \mathcal{T}_{h,\star}^\ell |$ on a slab
$n=\star$, for Sec. \ref{sec:3:examples}.}
\label{tab:1:convergence}
\end{table}
By Fig. \ref{fig:9:tauDistribution}
the distribution of the time subinterval lengths of $\mathcal{Q}_n^\ell$
over $I$ are illustrated.
The goal \eqref{eq:10:goalTol}
is reached for the 18th loop of the optimisation problem
(Tab. \ref{tab:1:convergence}) with an absolute error
$\| u - u_{\tau,h}^{18} \|_{\mathcal{Q}_c} = 5.744 \cdot 10^{-4}$.

The numerical results (Fig. \ref{fig:8:solution}-\ref{fig:9:tauDistribution})
highlight the local space-time refinement characteristic of the solver
in a self-adaptive way,
while the finer space-time cells are located inside and
close to the boundary of the control volume
$\mathcal{Q}_{c,\text{KB2d}}(\boldsymbol x, t)$, $t \in (0.25,1)$.
Remark that not any refinement is done for $t > 1$ of this parabolic
problem since the control volume is not active.


\section{Impact}
\label{sec:4:impact}

The software gives new and efficient insights to the implementation of
goal-oriented mesh adaptivity which supports the development of cost-effective
numerical screening tools based on finite element approaches for fundamental
problems in science and engineering.
Derivatives of the software have already been used for
stabilised stationary and instationary convection-dominated problems with
goal-oriented mesh adaptivity in \cite{Bruchhaeuser2018,Bruchhaeuser2018a}
as well as for coupled deformation and diffusion-driven transport in
\cite{Bause2017,Koecher2015}.
The intended user group of researchers and engineers using and developing
finite element based numerical simulation screening tools for several problems
is widespread and globally distributed; an example would be the extension of
the work \cite{Ahmed2015} in which they are using higher order variational time
discretisations for convection-dominated transport with non-goal oriented
adaptive time step control,
or, an efficient instationary extension to the approach presented in
\cite{Endtmayer2017}.
The software is currently not used in commercial settings,
but the license allows for the integration into commercial tools.

\section{Conclusions}
\label{sec:5:conclusions}

This original software publication provides
efficient and scalable data structures and algorithms
for the implementation of goal-oriented mesh adaptivity
in a highly modular way.
The key ideas of the data structures and algorithms can be reused,
even independently of the programming language,
in any adaptive finite element code.
The performance and applicability of the software is shown with an illustrative
example and several others are shipped with the code.
The work yields a major breakthrough
as a freely available open-source implementation
with extensive in-source documentation
for the dual weighted residual method in an application-free way
such that it can be easily adopted by other frameworks.
Ongoing work is on the distributed-memory parallelisation,
on the serialisation of the data storage hiding memory operations,
and on finding optimal tuning parameters of the minimisation problem.





\begin{thebibliography}{99}

\bibitem{Becker1998}
{\sc R.~Becker and R.~Rannacher}, %
{\em Weighted a posteriori error control in {F}{E} methods},
In Proc. of the 2nd EC on Numer. Math. and Adv. Appl. ENUMATH 1997,
H.G.~Bock et.al. (eds),
World Scientific, Singapore, 1998.

\bibitem{Becker2001}
{\sc R.~Becker and R.~Rannacher}, %
{\em An optimal control approach to a posteriori error estimation in finite element methods},
Acta Numer. 10:1-102,
\doi{10.1017/S0962492901000010}, 2001.

\bibitem{Endtmayer2017}
{\sc B.~Endtmayer and T.~Wick}, %
{\em A partition-of-unity dual-weighted residual approach for multi-objective
error estimation applied to elliptic problems},
Comput. Meth. Appl. Math. 17(4):1-25, \doi{10.1515/cmam-2017-0001}, 2017.

\bibitem{Bruchhaeuser2018}
{\sc M.P.~Bruchh{\"{a}}user, K.~Schwegler and M.~Bause}, %
{\em Numerical study of goal-oriented error control for stabilized finite element
methods},
In Adv. {F}inite {E}lement Meth. with Appl., T.~Apel et.al. (eds),
Lecture Notes in Comput. Sci. and Engrg., Springer, Berlin,
p.1-19, accepted, \arXiv{1803.10643}, 2018.

\bibitem{Bruchhaeuser2018a}
{\sc M.P.~Bruchh{\"{a}}user, K.~Schwegler and M.~Bause}, %
{\em Dual weighted residual based error control for nonstationary
convection-dominated equations: potential or ballast?},
submitted, p.1-13, \arXiv{1812.06810}, 2018.

\bibitem{Bangerth2003}
{\sc W.~Bangerth and R.~Rannacher}, %
{\em Adaptive {F}inite {E}lement {M}ethods for {D}ifferential {E}quations},
Birkh{\"{a}}user, Basel, 2003.

\bibitem{dealIIReferenceV90}
{\sc G.~Alzetta, D.~Arndt, W.~Bangerth, V.~Boddu, B.~Brands, D.~Davydov,
R.~Gassmoeller, T.~Heister, L.~Heltai, K.~Kormann, M.~Kronbichler, M.~Maier,
J.-P.~Pelteret, B.~Turcksin and D.~Wells}, %
{\em The deal.II library, Version 9.0},
J. Numer. Math. 26(4):173-183, \doi{10.1515/jnma-2018-0054}, 2018.

\bibitem{Koecher2015}
{\sc U.~K{\"{o}}cher}, %
{\em Variational space-time methods for the elastic wave equation and
the diffusion equation}, Ph.D. thesis,
Mech. Engrg. Helmut-Schmidt-University Hamburg,
p. 1-188, \urn{urn:nbn:de:gbv:705-opus-31129}, 2015.

\bibitem{Bause2017}
{\sc M.~Bause, F.A.~Radu and U.~K{\"{o}}cher}, %
{\em Space-time finite element approximation of the {B}iot poroelasticity system
with iterative coupling},
Comput. Meth. Appl. Mech. Engrg. 320:745-768, \doi{10.1016/j.cma.2017.03.017}, 2017.


\bibitem{Ahmed2015}
{\sc N.~Ahmed and V.~John}, %
{\em Adaptive time step control for higher order variational time discretizations
applied to convection-diffusion equations},
Comput. Meth. Appl. Mech. Engrg. 285:83-101, \doi{10.1016/j.cma.2014.10.054}, 2015.



\end{thebibliography}


\renewcommand\refname{\vskip-6ex} 

\newpage
\onecolumn
\section*{Required Metadata}

\section*{Current code version}

Ancillary data table required for subversion of the codebase.
Kindly replace examples in right column with the correct information about your
current code, and leave the left column as it is.

\begin{table}[!h]
\begin{tabular}{|l|p{6.5cm}|p{6.5cm}|}
\hline
\textbf{Nr.} & \textbf{Code metadata description} & \textbf{Please fill in this column} \\
\hline
C1 & Current code version & \texttt{Version 1.0.0} \\
\hline
C2 & Permanent link to code/repository used for this code version &
\texttt{https://github.com/} \texttt{dtm-project/dwr-diffusion} \\
\hline
C3 & Legal Code License   &
\texttt{https://github.com/} \texttt{dtm-project/dwr-diffusion/License} \\
\hline
C4 & Code versioning system used & \texttt{git} \\
\hline
C5 & Software code languages, tools, and services used &
\texttt{C++.17}; \texttt{cmake}, \texttt{gcc}, \texttt{mpi},
optional: \texttt{paraview}, \texttt{doxygen}\\
\hline
C6 & Compilation requirements, operating environments \& dependencies &
\texttt{deal.II} with \texttt{hdf5};
Linux (Fedora, CentOS 7, RHEL 7, etc.), MacOS, Windows WSL \\
\hline
C7 & If available Link to developer documentation/manual &
\texttt{https://github.com/} \texttt{dtm-project/dwr-diffusion} \\
\hline
C8 & Support email for questions & \texttt{dtmproject@uwe.koecher.cc} \\
\hline
\end{tabular}
\caption{Code metadata (mandatory)}
\label{tab:code:metadata} 
\end{table}

%
%

\end{document}